\documentclass[11pt,a4paper,useAMS,usenatbib]{emulateapj}
\usepackage{rotating}
\usepackage{epsfig}
\usepackage{amsmath}
\usepackage{natbib}
\bibpunct{(}{)}{;}{a}{}{,}

\usepackage{float}
\usepackage{color}

\begin{document}

\title{Detection of the missing baryons toward the sightline of H\,1821+643}

\author{Orsolya E. Kov\'acs\altaffilmark{1,2,3}, \'Akos Bogd\'an\altaffilmark{1}, \\ Randall K. Smith\altaffilmark{1}, Ralph P. Kraft\altaffilmark{1},  and William R. Forman\altaffilmark{1}}
\affil{\altaffilmark{1}Harvard Smithsonian Center for Astrophysics, 60 Garden Street, Cambridge, MA 02138, USA; orsolya.kovacs@cfa.harvard.edu}
\affil{\altaffilmark{2}Konkoly Observatory, MTA CSFK, H-1121 Budapest, Konkoly Thege M. {\'u}t 15-17, Hungary}
\affil{\altaffilmark{3}E{\"o}tv{\"o}s University, Department of Astronomy, Pf. 32, 1518, Budapest, Hungary}
\shorttitle{DETECTING THE MISSING BARYONS}
\shortauthors{KOV\'ACS ET AL.}

\begin{abstract}

Based on constraints from Big Bang nucleosynthesis and the cosmic microwave background, the baryon content of the high-redshift Universe can be precisely determined.
However, at low redshift, about one-third of the baryons remain unaccounted for, which poses the long-standing \textit{missing} baryon problem.
The missing baryons are believed to reside in large-scale filaments in the form of warm-hot intergalactic medium (WHIM). In this work, we employ a novel stacking approach to explore the hot phases of the WHIM.
Specifically, we utilize the $470$~ks \textit{Chandra} LETG data of the luminous quasar, H\,1821+643, along with previous measurements of UV absorption line systems and spectroscopic redshift measurements of galaxies toward the quasar's sightline.
We repeatedly blueshift and stack the X-ray spectrum of the quasar corresponding to the redshifts of the 17 absorption line systems.
Thus, we obtain a stacked spectrum with $8.0$ Ms total exposure, which allows us to probe X-ray absorption lines with unparalleled sensitivity.
Based on the stacked data, we detect an O\,VII absorption line that exhibits a Gaussian line profile and is statistically significant at the $3.3 \, \sigma$ level.
Since the redshifts of the UV absorption line systems were known a priori, this is the first definitive detection of an X-ray absorption line originating from the WHIM.
The equivalent width of the O\,VII line is $(4.1\pm1.3) \ \mathrm{m\AA}$, which corresponds to {an O\,VII} column density of $(1.4\pm0.4)\times10^{15} \ \mathrm{cm^{-2}}$.
{We constrain the absorbing gas to have a density of $n_{\rm H} = (1-2)\times10^{-6} \ \rm{cm^{-3}}$ for a single WHIM filament.}
{We derive $\Omega_{\rm b} \rm(O\,VII) = (0.0023 \pm 0.0007)   \, \left[ f_{O\,VII} \, {Z/Z_{\odot}} \right]^{-1}$ for the cosmological mass density of O\,VII, assuming that all 17 systems contribute equally. }

\end{abstract}


\section{Introduction}
\label{sec:intro}

The mass density of baryons in the high-redshift Universe can be determined from Big Bang nucleosynthesis and from the cosmic microwave background. These independent methods, in agreement with each other, have established that the cosmic baryonic mass density is $\Omega_{\rm b} \approx 0.0449$ \citep[e.g.][]{2014ApJ...781...31C,2016A&A...594A..13P}. Similar, but slightly lower values are obtained from Lyman-alpha (Ly-$\rm{\alpha}$) studies at $z>2$ \citep[e.g.][]{2005PhRvD..71j3515S}. However, about one-third of the baryons are unaccounted for in the low-redshift ($z<2$) universe compared {to observations and theoretical predictions} \citep{2001ApJ...552..473D,2002ApJ...564..604F,2011ApJ...731....6S,2012ApJ...759...23S}. This discrepancy exists in both individual galaxies and in the large-scale structure universe, introducing the \textit{local} and \textit{global} missing baryon problem.

According to the local missing baryon problem, the baryon fraction of galaxy halos falls short of the cosmic value \citep{2007ARA&A..45..221B,2011ApJ...737...22A,2013ApJ...772...97B}. The existence of  hot gaseous halos around galaxies is a fundamental prediction of all galaxy formation models \citep{1978MNRAS.183..341W,1991ApJ...379...52W,2002MNRAS.335..799T,2010MNRAS.407.1403C} and theoretical studies suggest that reservoirs of large amounts of hot gas in the dark matter halos of galaxies may resolve the local missing baryon problem. Given the momentous nature of this question, a wide range of studies have attempted to study the hot X-ray halos of galaxies. The gaseous halos of massive spiral galaxies were recently explored using \textit{Chandra} and \textit{XMM-Newton} observations \citep{2011ApJ...737...22A,2013ApJ...772...98B,2013ApJ...772...97B,2016MNRAS.455..227A,2017ApJ...850...98B,2018ApJ...855L..24L}. However, these observations were only able to characterize the gaseous halos out to $60$\,kpc (${\sim} 0.15\,\mathrm{r_{200}}$ )\footnote{$r_{200}$ corresponds to the radius where the average matter density is $\geq$\,$200$ times the critical density of the universe.}. Since only $1{-}3 \%$ of the hot gas resides within this volume, approximately $97{-}99 \%$ of the gas remains undetected.

Globally, the missing baryons are believed to be in the form of warm-hot intergalactic medium (WHIM). The WHIM is a low-density, shock-heated medium that is concentrated in large-scale, filamentary structures that follow the dark matter distribution and connect virialized structures \citep[e.g.][]{1987ApJ...313..505W,2005MNRAS.359..272C,2008MNRAS.383.1655S}. Collapsed objects, such as galaxies, galaxy groups/clusters, contain only ${\sim}18$\,\% of the baryons \citep{1992MNRAS.258P..14P,2012ApJ...759...23S}, and the dominant fraction of baryons is expected to reside in the warm ($T \lesssim 10^5$~K) and hot ($T\gtrsim 10^5$~K) phases of the WHIM. The existence of the warm phases of the WHIM has been confirmed by absorption studies in the UV regime. These observations have explored the most abundant ions, such as O\,VI, Ne\,VIII and H\,I Ly-$\rm{\alpha}$, using various AGN sightlines \citep{2001ApJ...559L...5C,2005ApJ...624..555D,2008ApJ...679..194D}. However, the {explored} warm phases of the WHIM account only for $20-30$\% of the baryon content of the universe \citep{2012ApJ...759...23S}. Although it is hypothesized that a notable fraction of the missing baryons lie in the hot phases of the WHIM, the conclusive observational evidence is still lacking \citep{2007ARA&A..45..221B}. While the densest and hottest parts of the WHIM have been detected in X-ray emission studies \citep{2008A&A...482L..29W,2016ApJ...818..131B,2018ApJ...858...44A}, it is expected that the dominant fraction of the WHIM resides at low densities ($n_{\rm H} \lesssim 10^{-5} $) that are not accessible by current emission studies.

To resolve the missing baryon problem, the higher ionization states of heavy elements must be studied, such as O\,VII and O\,VIII, whose absorption transitions are at X-ray wavelengths. X-ray absorption studies provide the ideal framework for these investigations. Specifically, the gaseous halos of galaxies and the large-scale WHIM can imprint their spectral signatures (i.e.\ absorption lines) on the observed spectrum of background quasars. Although the robustness of this method has been demonstrated by UV absorption studies, X-ray absorption studies are hindered by the low effective area of \textit{Chandra} and \textit{XMM-Newton} gratings. For example, the Cosmic Origin Spectrograph aboard \textit{HST}, which is routinely used to detect UV absorption lines, has an effective area of $A = 2000 \ \rm{cm^{2}}$ and a resolution of $R=20000$, which surpasses \textit{Chandra} ACIS-LETG, which has $A \approx 17 \ \rm{cm^{2}}$ and $R \approx 430$ at the wavelength of the O\,VII line ($\lambda = 21.6 \ \rm{\AA}$)\footnote{These values apply to \textit{Chandra} Cycle 2, when the H\,1821+643 observations were taken.}. 

In the past few decades, several authors claim to have detected WHIM filaments. Notably, \citet{2005Natur.433..495N} reported two individual O\,VII absorbers in the spectrum of Mkn\,421 using \textit{Chandra} ACIS-LETG observations. However, these detections remain controversial. First, the results were not reproduced with the \textit{XMM-Newton} RGS, althouggh it is also unclear if the resolution of RGS is sufficient to confirm or reject the detections \citep{2006ApJ...642L..95W}. Second, \citet{2005Natur.433..495N} did not take into account that the detected absorption lines were obtained in a blind search, and hence the significance of the detection may be overestimated \citep{2006ApJ...652..189K}.
Recently, \citet{2018Natur.558..406N} carried out a blind search toward the sightline of 1ES\,1553+113 using \textit{XMM-Newton} RGS observations. They reported two O\,VII absorbers and associated them with the WHIM. 
However, the observed O\,VII equivalent widths ($\gtrsim10 \, \mathrm{m\AA}$) are typical for inner galactic halos, while O\,VII equivalent widths for WHIM filaments are expected to be about an order of magnitude lower \citep{2017SPIE10397E..0QS}.
In addition, weak absorption lines with a statistical significance of ${\sim} 2\sigma$ were detected in the spectrum of H1821+643 \citep{2003ApJ...582...82M}. Given the low statistical significance of these detections \citep{2006ApJ...652..189K}, these results are not definitive. Thus, nearly 20 years after the launch of the current generation of X-ray telescopes, observational constraints on the baryon content of the hot phase of the WHIM remain elusive.

The present study is the first in a series of papers that aim to address the missing baryon problem using a stacking approach. Specifically, we utilize \textit{Chandra} LETG spectra of luminous quasars along with a priori redshift measurements of UV (Ly-$\rm{\alpha}$ and O\,VI) absorption line systems and we co-add (i.e. stack) the X-ray line forest from each absorption line system for major X-ray metal lines. Given the multitude of UV absorption systems in the sightline of quasars, this method increases the effective exposure time by the number of the absorption line systems. Thus, we can statistically probe the low-density gas associated with the WHIM to unprecedentedly sensitive limits, thereby circumventing the low effective area of X-ray grating instruments.

\begin{table}
\caption{The analyzed \textit{Chandra} observations}
\begin{minipage}{8.5cm}
\renewcommand{\arraystretch}{1.3}
\centering
\begin{tabular}{c c c c}
\hline
Obs ID& $t_{\rm exp}$ [ks] &Instrument &Date \\
\hline
2186	& 165.1	& ACIS-LETG	& 2001 Jan 18	\\
2310	& 163.2	& ACIS-LETG	& 2001 Jan 21	\\
2311	& 90.1	& ACIS-LETG	& 2001 Jan 24	\\
2418	& 51.8	& ACIS-LETG	& 2001 Jan 17	\\
\hline \\
\end{tabular}
\end{minipage}
\label{tab:obs}
\end{table}

We note that a similar method was used by \citet{2010ApJ...716.1514Y}, who stacked \textit{Chandra} LETG spectra using the redshifts of previously detected O\,VI absorption line systems. Given that simulations suggest coincidences between the presence of O\,VI and O\,VII lines \citep{2012ApJ...753...17C}, this is a promising approach. However, UV absorption studies detect significantly fewer O\,VI than Ly-$\rm{\alpha}$ absorbers. Hence, the sensitivity of this method did not result in statistically significant detections. Including \mbox{Ly-$\rm{\alpha}$} absorbers can drastically increase the signal-to-noise ratios. Since Ly-$\rm{\alpha}$ absorption line systems may be associated with low-mass galaxies residing in low-density voids or may be detached from galaxies entirely \citep{2014MNRAS.441.2923C,2017ApJ...837...29R}, it is best to utilize follow-up optical spectroscopy of the galaxies in the field-of-view of the background quasar. This, in turn, allows the redshifts of the absorption line systems to be cross-correlated with those of the galaxies. For the stacking analysis, we utilize only those absorption line systems that are associated with relatively massive galaxies ($M_{\mathrm{halo}} \gtrsim 3 \times 10^{11} \rm \, M_{\sun}$), thereby increasing the chance of stacking the filaments of the WHIM because the more massive galaxies are likely to lie in filaments.

\begin{table*}
\caption{The list of Ly-$\rm{\alpha}$ absorption lines used in this study}
\begin{minipage}{18cm}
\renewcommand{\arraystretch}{1.3}
\centering
\begin{tabular}{c c c c c c c c c}
\hline
$z_{\rm abs}$ & $z_{\rm gal}$ &  R.A.   &  Decl.  & $M_{\rm \star}$  & $M_{\rm halo}$ & $R_{\rm vir}$ & $b$ & $ R_{\rm vir}/b$   \\
 & & (J2000)& (J2000) & [$\rm{10^{10} \ M_{\odot}}$] & [$10^{12} \ \rm{M_{\odot}}$] & [kpc] & [kpc] & \\
(1) & (2) & (3) & (4) & (5) & (6) & (7) & (8) & (9) \\
\hline
0.05704  & 0.05683	& 18 21 21.38 & 64 49 40.4	& 0.95	& 0.35	&143.2	&1911.8	&	0.07\\
0.06432  & 0.06553	& 18 16 31.37 & 64 43 54.6	& 18.20	& 73.07	&839.4	&3122.8	&	0.27\\
0.08910  & 0.08930	& 18 20 42.96 & 64 19 45.8	& 3.39	& 1.07	&207.1	&791.7	&	0.26\\
0.11152  & 0.11155	& 18 20 53.45 & 64 19 37.1	& 3.99	& 1.36	&224.5	&828.4	&	0.27\\
0.11974  & 0.11957	& 18 25 04.45   & 64 25 22.9	& 3.55	& 1.14	&211.7	&2619.2	&	0.08\\
0.12157  & 0.12154	& 18 22 02.65  & 64 21 39.3	& 1.82	& 0.54	&165.6	&154.3	&	1.07\\
0.12385  & 0.12258	& 18 23 58.19 & 64 26 52.3	& 21.38	& 132.25	&1022.7	&1863.7	&	0.55\\
0.14760  & 0.15039	& 18 21 08.73   & 63 52 09.0	& 17.38	& 62.07	&795.1	&4405.5	&	0.18\\
0.16990  & 0.17086	& 18 21 36.61 & 64 21 25.0	& 4.37	& 1.59	&236.2	&399.7	&	0.59\\
0.18049  & 0.18012	& 18 19 55.22 & 64 35 08.4	& 6.92	& 4.10	&322.8	&3443.0	&	0.09\\
0.19905  & 0.20057	& 18 22 14.04 & 64 03 05.4	& 26.92	& 329.44	&1386.0	&3357.8	&	0.41\\
0.22489  & 0.22560	& 18 21 54.40 & 64 20 09.3	& 2.40	& 0.70	&180.3	&112.5	&	1.60\\
0.24132  & 0.24435	& 18 21 56.28 & 64 22 50.4	& 3.89	& 1.31	&221.8	&495.1	&	0.45\\
0.24514  & 0.24568	& 18 22 19.11 & 64 18 43.5	& 1.10	& 0.38	& 147.1	&669.6	&	0.22\\
0.25814  & 0.25147	& 18 20 31.50 & 64 20 24.0	& 4.90	& 1.96	&253.0	&2089.1	&	0.12\\
0.26156  & 0.26650	& 18 21 00.73   & 64 37 54.3	& 10.00	& 10.73	& 443.9	&4032.9	&	0.11\\
0.26660  & 0.26669	& 18 22 10.26 & 64 17 15.6	& 10.47	& 12.25	& 463.8	&851.8	&	0.54\\
\hline \\
\end{tabular}
\end{minipage}
Columns are as follows: (1) Redshift of the Ly-$\rm{\alpha}$ absorption line system \citep{1998ApJ...508..200T}; (2) Redshift of the foreground galaxy associated with the UV absorption line taken from \citet{1998ApJ...508..200T}; (3) and (4) Right ascension and declination of the foreground galaxy; (5) Stellar mass of the galaxy derived from SDSS broadband photometric data and the FAST code; (6) Halo mass of the galaxy derived from the stellar mass and the stellar mass-halo mass relation of \citet{2010ApJ...717..379B}; (7) Virial radius of the galaxy; (8) Impact parameter of the galaxy at the redshift of the absorption line system; (9) The ratio of impact parameter and virial radius.
\label{tab:abs}
\end{table*}

In this work, we address the missing baryon problem using \textit{Chandra} ACIS-LETG observations (Table~\ref{tab:obs}) and previously detected UV absorption line systems toward the sightline of H\,1821+643. This {X-ray} luminous  ($F_{X}= 9.66 \times 10^{-12} \ \rm{erg \ s \ cm^{-2}}$) quasar resides at $z=0.297$, which makes it ideal to study absorption line systems along its sightline \citep{2015JATIS...1d5003B}. Indeed, the UV absorption line systems and the redshifts of galaxies around H\,1821+643 were extensively studied by \citet{1998ApJ...508..200T}. These authors identified 35 absorption line systems that may be associated with galaxies, whose redshifts are used for our stacking analysis. 

This paper is structured as follows. In Section \ref{sec:analysis} we introduce the analyzed set of UV absorption lines and we discuss the \textit{Chandra} observations and their analysis. In Section \ref{sec:results}, we present the stacked spectrum of H\,1821+643 and study the detected O\,VII absorption line. In Section \ref{sec:disc}, we discuss our results. Finally, a summary is given in Section \ref{sec:conclusion}. For cosmological parameters, we assume $H_{\rm{0}}=70 \ \rm{km \ s^{-1} \ Mpc^{-1}}$, $ \Omega_M=0.3$, and $\Omega_{\Lambda}=0.7$, and all error bars represent $1\sigma$ uncertainties.

\begin{figure*}[!]
\begin{center}
\leavevmode
	\epsfxsize=0.8\textwidth \epsfbox{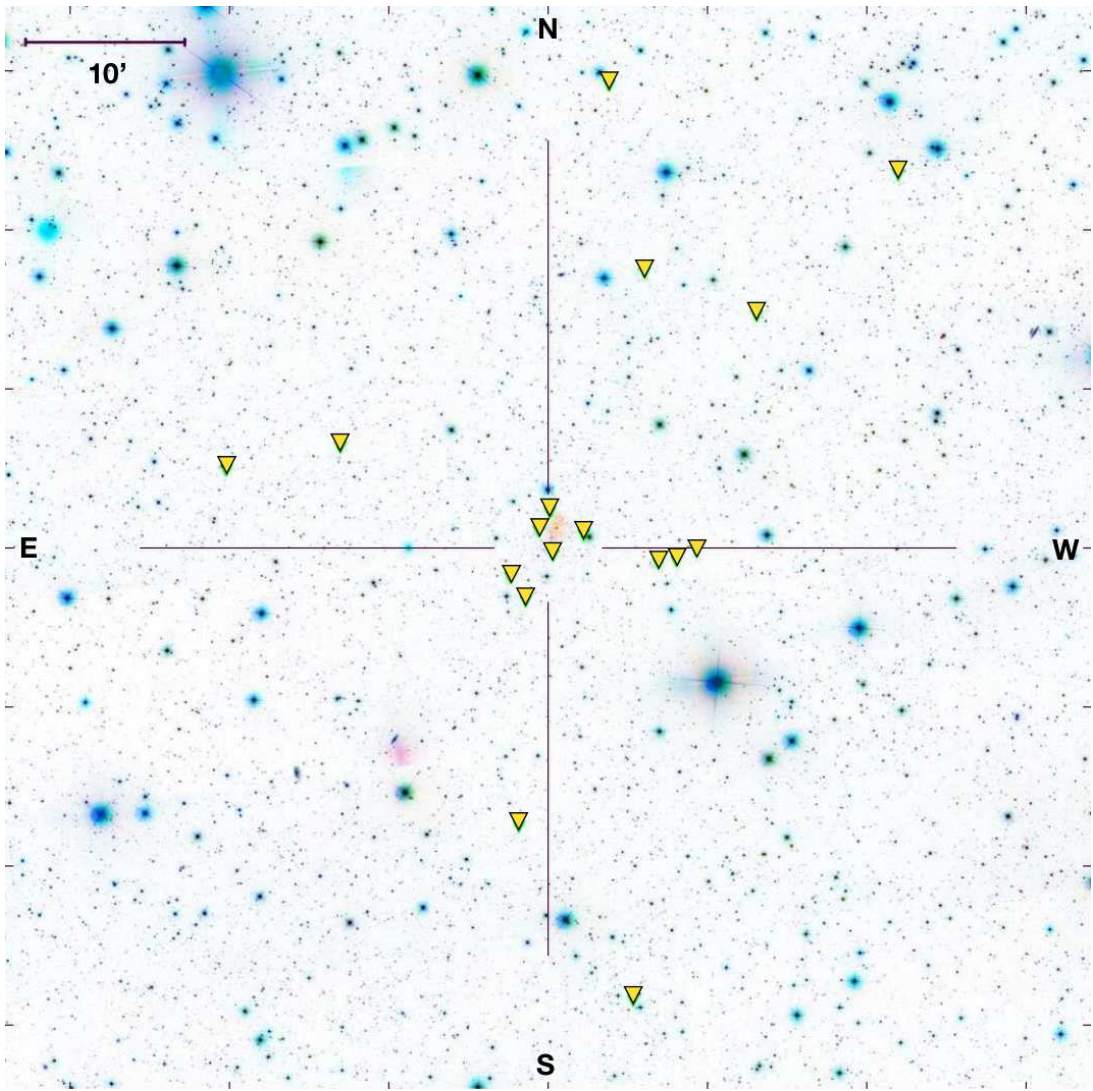}
	\vspace{0cm}
	\caption{SDSS \textit{u}-band image of the $1\degr.14 \times 1\degr.14$ environment of the luminous AGN, H\,1821+643, which is located at the center of the image. The downward pointing symbols mark the positions of the 17 galaxies that are associated with Ly-$\rm{\alpha}$ absorbers \citep[see][and Section \ref{sec:uv}]{1998ApJ...508..200T}. {Note that the galaxies are distributed across the entire field-of-view and are not strongly clustered.} We used the redshifts of the 17 absorption line systems to stack the \textit{Chandra} ACIS-LETG spectrum of H\,1821+643. Given the distant nature of the foreground galaxies, they are {not easily visible} in the SDSS image. Note that the foreground galaxies are not clustered and they cover a wide range in impact parameter in the range of $b=0.1-4.1$ Mpc (see Table~\ref{tab:abs}). North is up and east is on the left of the image.}
	\label{fig:galdistr}
\end{center}
\end{figure*}

\section{Data analysis}
\label{sec:analysis}

\subsection{UV absorption line systems}
\label{sec:uv}

Based on high-resolution UV data taken with the Hubble Space Telescope, \citet{1998ApJ...508..200T} studied the Ly-$\rm{\alpha}$ absorbers in the spectrum of H\,1821+643. These authors also measured spectroscopic redshifts of 154 galaxies in the ${\sim}1\degr$ field centered on the AGN. Therefore, the redshifts of the UV absorption line systems and the intervening galaxies could be cross-correlated. Based on this, \citet{1998ApJ...508..200T} identified 35 absorption line systems that may be associated with foreground galaxies. Since the redshifts of several Ly-$\rm{\alpha}$ absorbers are similar, in some cases the same foreground galaxy is assigned to multiple absorption line systems. For the foreground galaxies with multiple absorption line systems, we retained one Ly-$\rm{\alpha}$ absorber, whose redshift was the closest to that of the foreground galaxy. This {reduced} our sample to 21 absorption line systems.

The mass distribution of galaxies is different in voids and in filaments. Specifically,  almost all massive galaxies (with halo mass of $M_{\rm halo} \gtrsim 3\times 10^{11} \ \rm{M_{\odot}}$) reside in filaments, while walls and voids are dominated by low-mass ($M_{\rm halo} \lesssim 3\times 10^{11} \ \rm{M_{\odot}}$) galaxies \citep{2007ApJ...655L...5A,2007MNRAS.375..489H,2014MNRAS.441.2923C}. 
To ensure that the Ly-$\rm{\alpha}$ absorbers are associated with filaments, we  further filtered the foreground galaxies based on their mass. 
To derive the stellar mass of the foreground galaxies, we used the \texttt{FAST} (Fitting and Assessment of Synthetic Templates) code that fits stellar population synthesis templates to broadband photometry data \citep{2009ApJ...700..221K}. We collected SDSS photometric data in the \textsl{u, g, r, i, z} bands for each foreground galaxy and we used the spectroscopic redshift of the galaxies as an input. For our model, we assumed solar metallicity, a Salpeter initial mass function \citep{1955ApJ...121..161S}, and a Milky Way dust law for the exctinction \citep{1989IAUS..135P...5C}.
To this end, we only retained galaxies whose stellar mass exceeds $M_{\star} \gtrsim 10^{10} \ \rm{M_{\odot}}$, which approximately corresponds to halo mass of $M_{\rm halo} \gtrsim 3 \times 10^{11} \ \rm{M_{\odot}}$ according to the stellar mass--halo mass relation of \citet{2010ApJ...717..379B}.

Based on the mass selection, we find that our sample consists of 17 absorption line systems, which are likely associated with massive ($M_{\star} \gtrsim 10^{10} \ \rm{M_{\odot}}$) foreground galaxies. In Table \ref{tab:abs} we show the redshift of the absorption line systems and the characteristics of the associated galaxies{, including impact parameter ($b$) and virial radius ($R_{\rm{vir}}$)}. We emphasize that the impact parameter of almost all foreground galaxies exceeds their virial radius, which implies that the present set of absorption line systems probe WHIM filaments rather than the halos of galaxies.  {We also note that the average Ly-$\rm{\alpha}$ equivalent width for our sample is $\sim 174.4$\,$\rm m\AA$. Thus, our stacking mostly concentrates on relatively strong Ly-$\rm{\alpha}$ absorbers.}

\begin{figure*}[!]
\begin{center}
\leavevmode
	\epsfxsize=0.8\textwidth \epsfbox{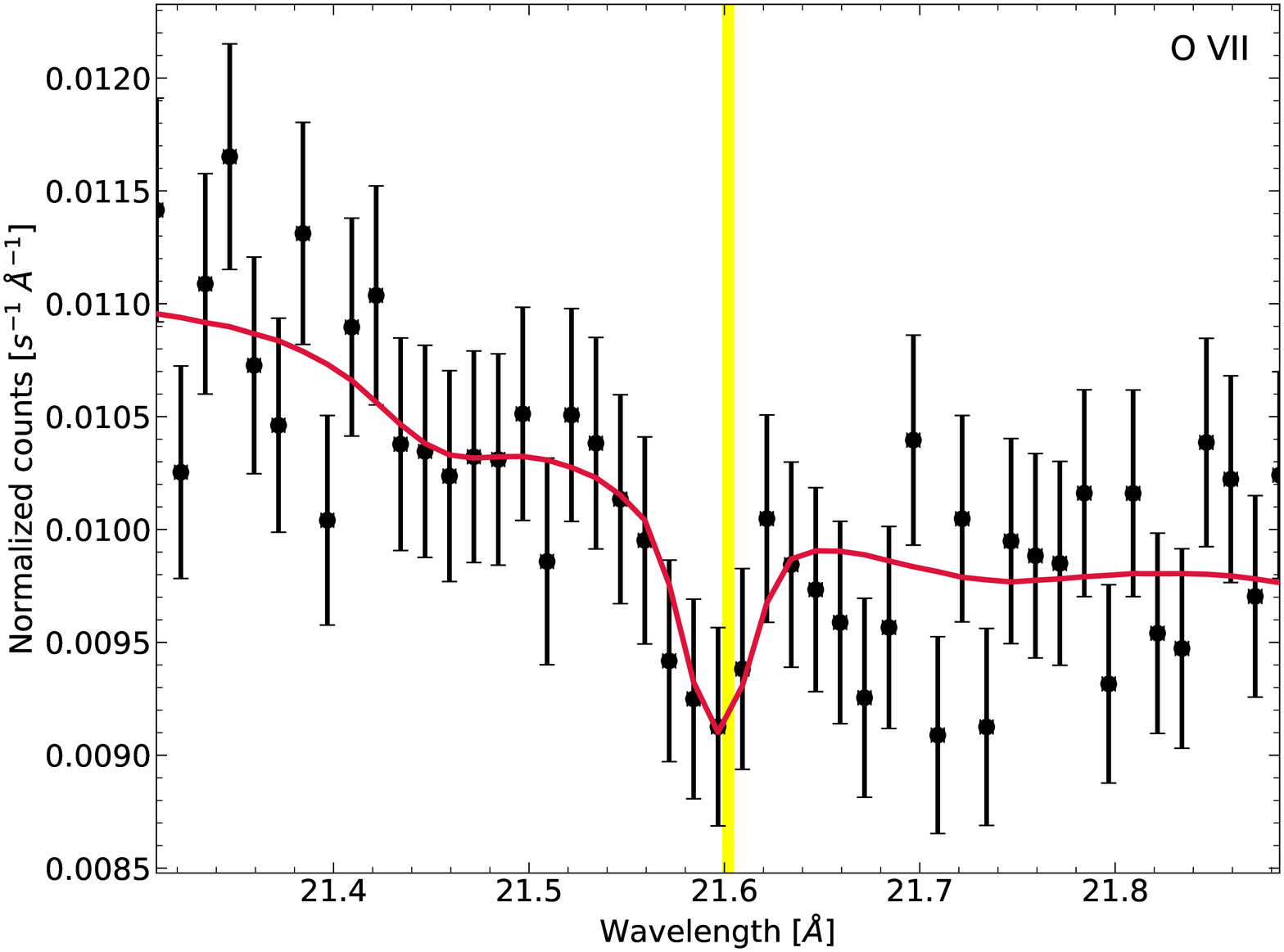}
	\vspace{0cm}
	\caption{Stacked \textit{Chandra} ACIS-LETG spectrum of H\,1821+643 around the rest-frame wavelength of the O\,VII ion. The spectrum is binned at $0.0125 \ \rm{\AA}$. To construct the stacked spectrum, we blueshifted the original $470$~ks spectrum 17 times corresponding to the redshift of each of the absorption line systems \citep{1998ApJ...508..200T}. The resulting spectra were then co-added. We stress that the redshifts of the absorption line systems were known a priori. Hence, this work does not involve a blind search for metal absorption lines. The combined exposure time of the stacked \textit{Chandra} spectrum is $8.0$\,Ms. At the rest-frame wavelength of O\,VII ($21.6 \ \rm{\AA}$), we detect an absorption line with a Gaussian line profile. The yellow vertical line denotes the rest-frame wavelength of O\,VII, red solid curve is the fitted model. The statistical significance of the detection is $3.3\sigma$. The corresponding equivalent width of the line is $(4.1\pm1.3) \ \rm{m\AA}$.} 
	\label{fig:h1821oline}
\end{center}
\end{figure*}

The resolving power of the \textit{Chandra} LETG is significantly lower than that of modern UV spectrographs. As such, the detected UV absorption lines may be spaced so close together in redshift space that they cannot be resolved by the LETG. This would result in stacking the same X-ray spectrum multiple times, which could distort our results. Therefore, a stacking study must avoid any overlaps by filtering \textit{duplicate} redshifts. We define overlapping UV absorption lines as those where the wavelengths corresponding to their redshifts are within $\delta \lambda = 0.0125  \ \rm{\AA}$, which is equivalent with one resolution element of \textit{Chandra} ACIS-LETG. For these systems, we would compute the mean redshifts of the overlapping systems and only include this mean redshift in the stacks. However, in our sample, we did not find overlapping redshifts. Therefore, we carry out our stacking analysis using the redshift of 17 Ly-$\rm{\alpha}$ absorber. The positions of the foreground galaxies, relative to H\,1821+643, are shown in Figure \ref{fig:galdistr}.

\subsection{\normalfont{Chandra} \textit{observations}}
\label{sec:data}

In this work, we analyze the \textit{Chandra} ACIS-LETG observations of H\,1821+643. This instrument/grating configuration is ideal to search for low-energy absorption lines. The AGN was observed in four pointings and the total exposure time of the data is $t_{\rm exp} = 470.2$\,ks. Details about the observations are provided in Table \ref{tab:obs}.  To prepare the data for the analysis, we rely on standard CIAO (version 4.9 and CALDB 4.7.6) software package tools \citep{2006SPIE.6270E..1VF}.

The main steps of the analysis are as follows. First, we reprocessed the data using the \texttt{chandra\_repro} {CIAO} tool to ensure that the most up-to-date calibration is applied. In addition, this tool creates the region masks, extracts the spectra, and builds the corresponding response files. In this work, we only considered the first order dispersed spectra and we combined the $\pm1$ orders of the grating spectra. Since H\,1821+643 was observed four times, we combined the individual observations using the \texttt{combine\_grating\_spectra} {CIAO} tool. This resulted in a single spectrum and the corresponding response files.

Because we stack the X-ray spectrum of H\,1821+643 multiple times corresponding to the redshifts of the UV absorption line systems, the spectra and the corresponding response matrix file (RMF) and ancillary response file (ARF) must be shifted. Specifically, we blueshifted the spectrum and the response files using $\lambda_{rest} = \lambda_{obs}(1+z)^{-1}$, where $z$ is the redshift of the individual absorption lines. This, in turn, shifts the lines to their rest-frame wavelengths, allowing us to stack the spectra associated with different UV absorption line systems. 

To stack the spectra and response files that are shifted to the rest-frame wavelength, it is necessary to have them on the same wavelength grid. However, the LETG wavelength grid is non-uniform, and hence the blueshifted spectra and response files cannot be directly co-added. To overcome this issue, we  applied two operations on the spectra and response files: rebinning and cropping. 

To rebin the data, we first defined a universal wavelength grid, which consists of bins with uniform widths of $0.0125 \, \rm{\AA} $. The particular choice was motivated by the fact that this bin size corresponds to a factor of four oversampling of the $0.05 \, \rm{\AA} $ resolution of LETG, and represents the default bin size of LETG. Therefore, the rebinned spectra have approximately the same number of bins as the original spectra, which means that rebinning does not distort any of the observed spectral features. To further confirm this, we rebinned the spectra and the response files using different binning factors, such as  $0.025 \, \rm{\AA} $ or $0.05 \, \rm{\AA} $. We conclude that the results presented in this work are not affected in any statistically significant way by the particular rebinning factor. 

Since the relevant metal lines are in the  $\lambda = 10-35 \, \rm{\AA}$ wavelength range, we cropped the spectra and retained only this wavelength range. Following these steps, we stacked the 17 blueshifted spectra and response files using \texttt{combine\_grating\_spectra} task.  

The end result of our stacking procedure is a single co-added blueshifted spectrum of H\,1821+643. This spectrum was used to search for absorption lines originating  from various metal lines. Given that we stacked the spectrum $N_{\rm abs} = 17$ times,  the total exposure time of our stacked spectrum is $t_{\rm stack} = t_{\rm exp} \times N_{\rm abs} = 8.0$\,Ms.

\section{Results} 
\label{sec:results}

\begin{figure*}
\begin{center}
\leavevmode
	\epsfxsize=0.48\textwidth \epsfbox{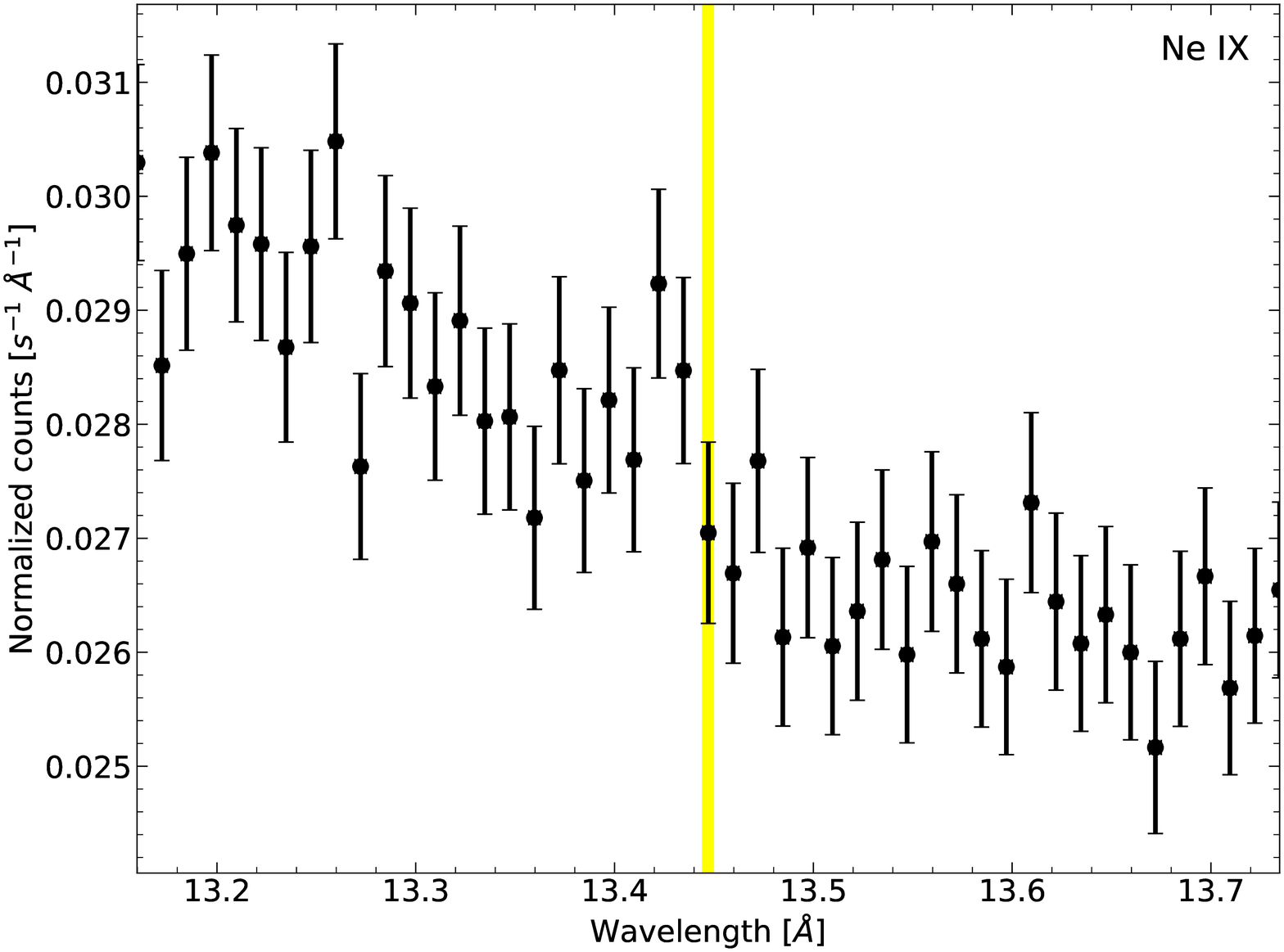}
	\epsfxsize=0.48\textwidth \epsfbox{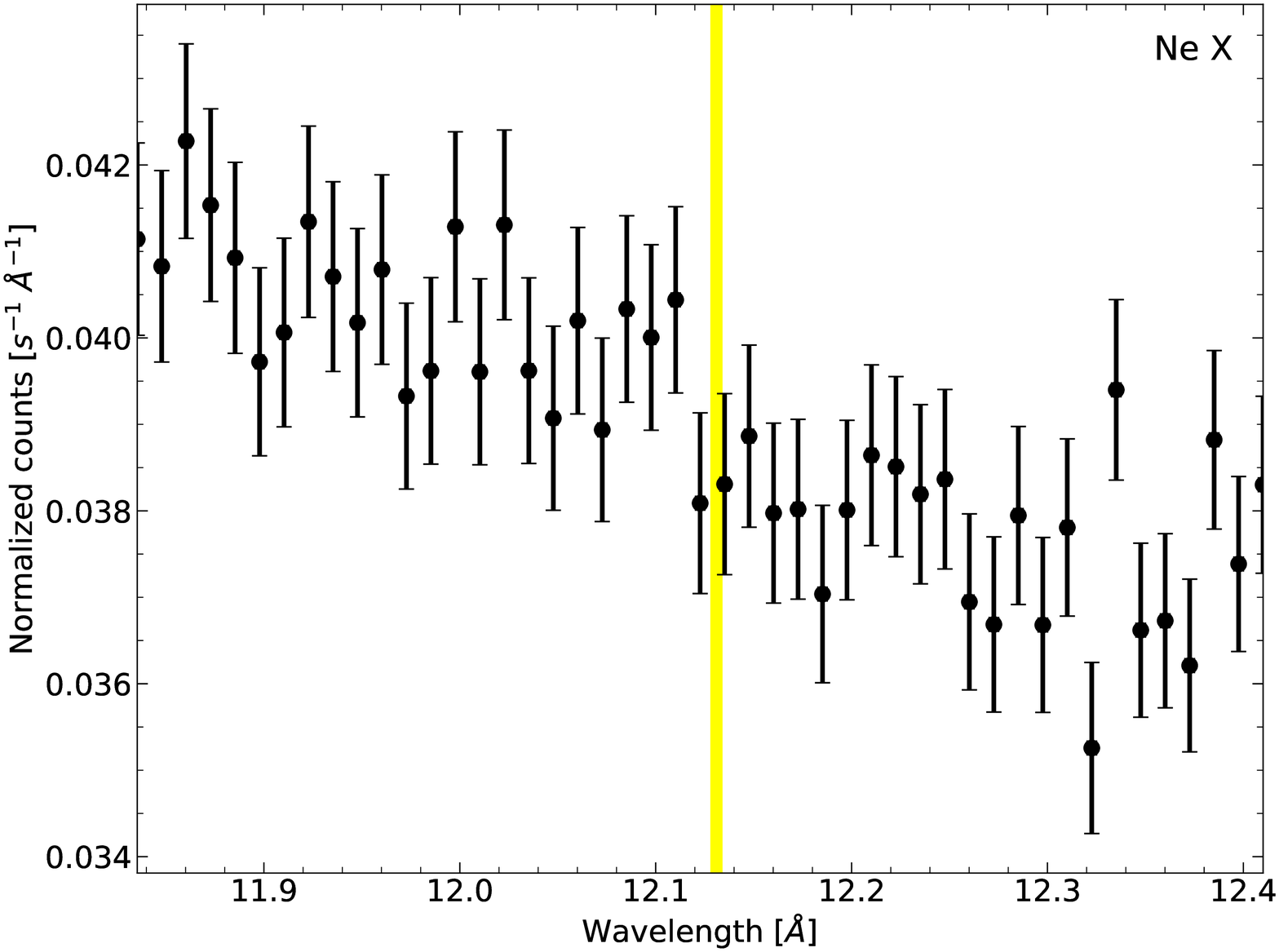}
	\epsfxsize=0.48\textwidth \epsfbox{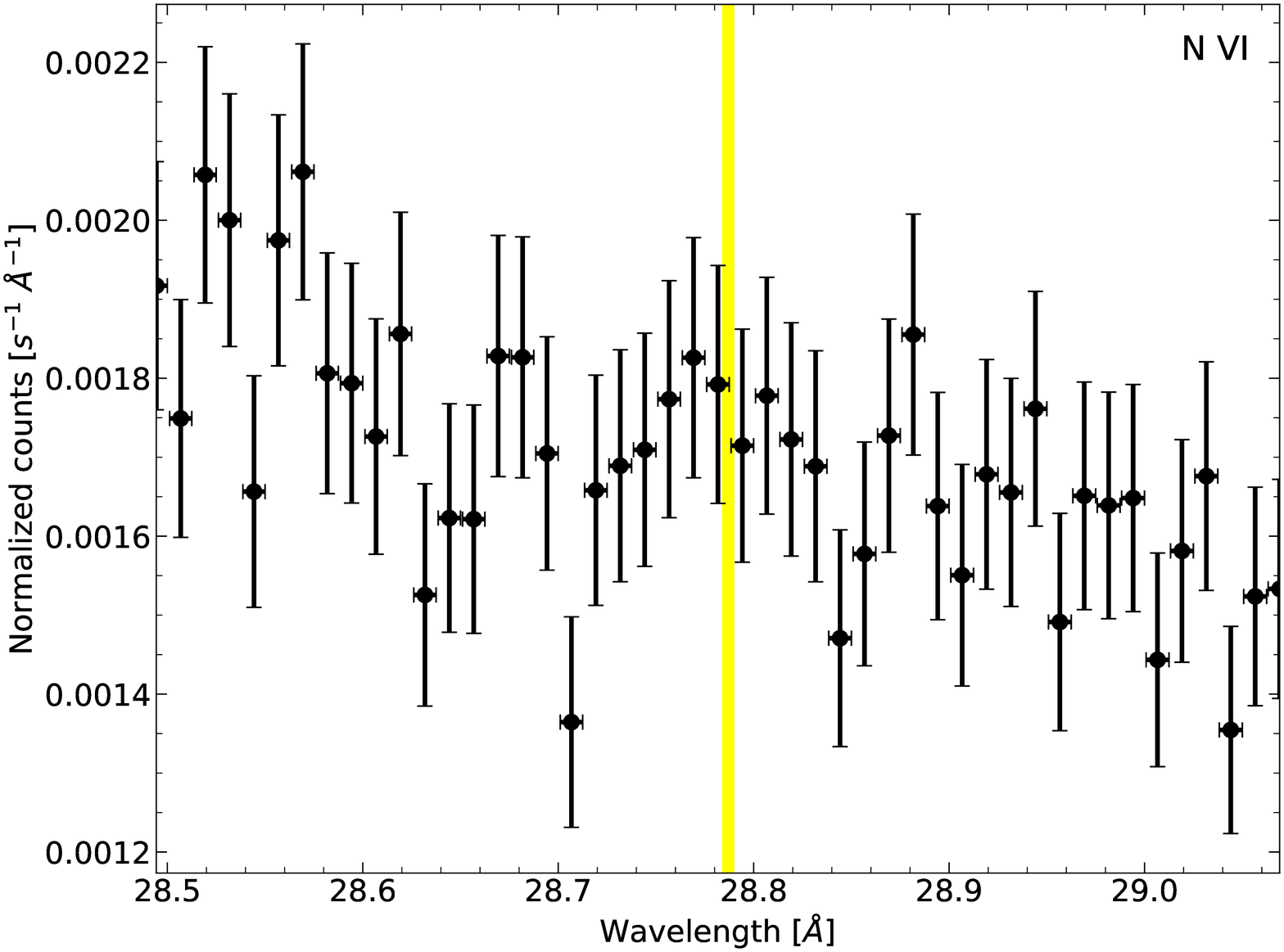}
	\epsfxsize=0.48\textwidth \epsfbox{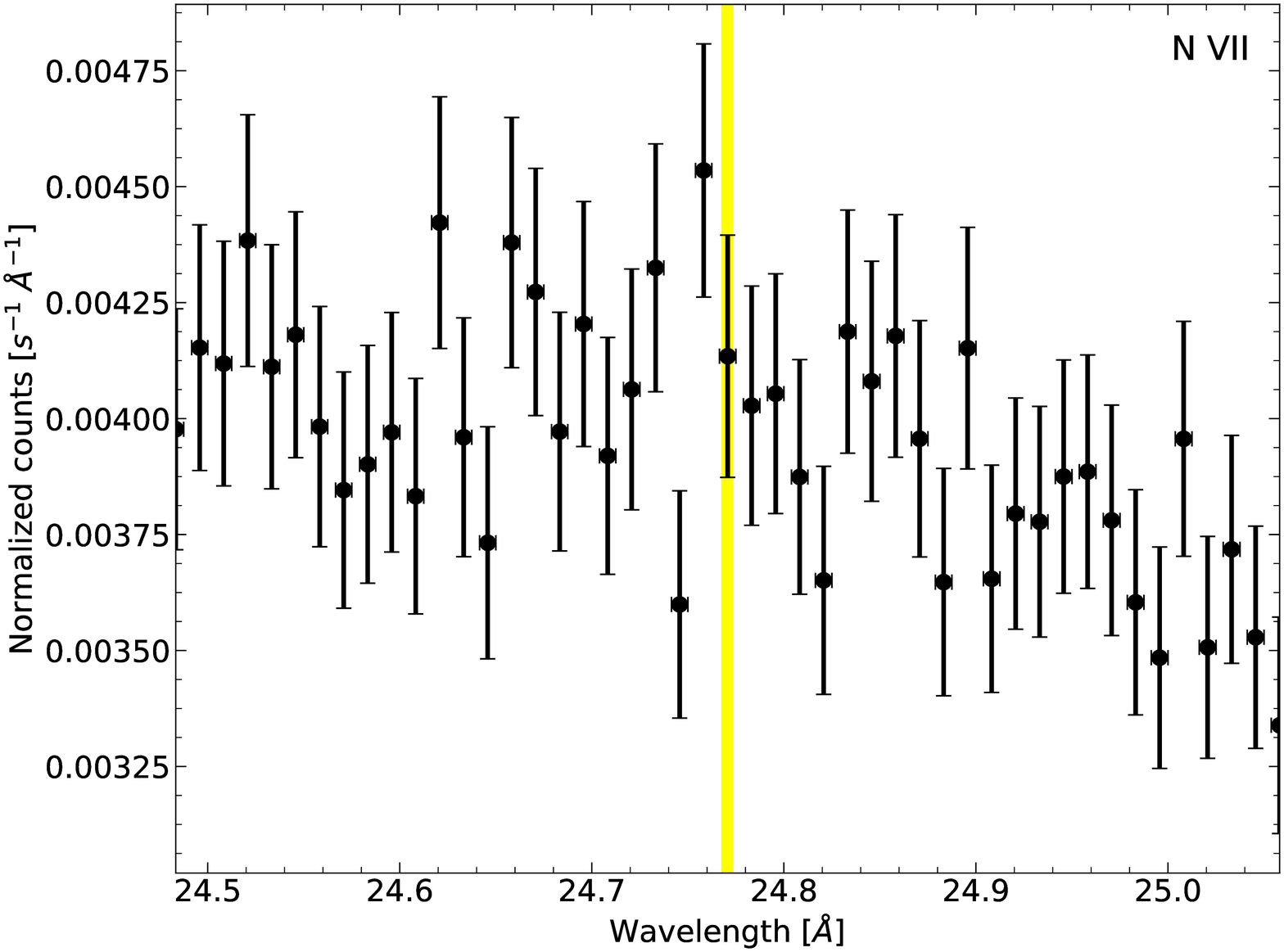}
	\epsfxsize=0.48\textwidth \epsfbox{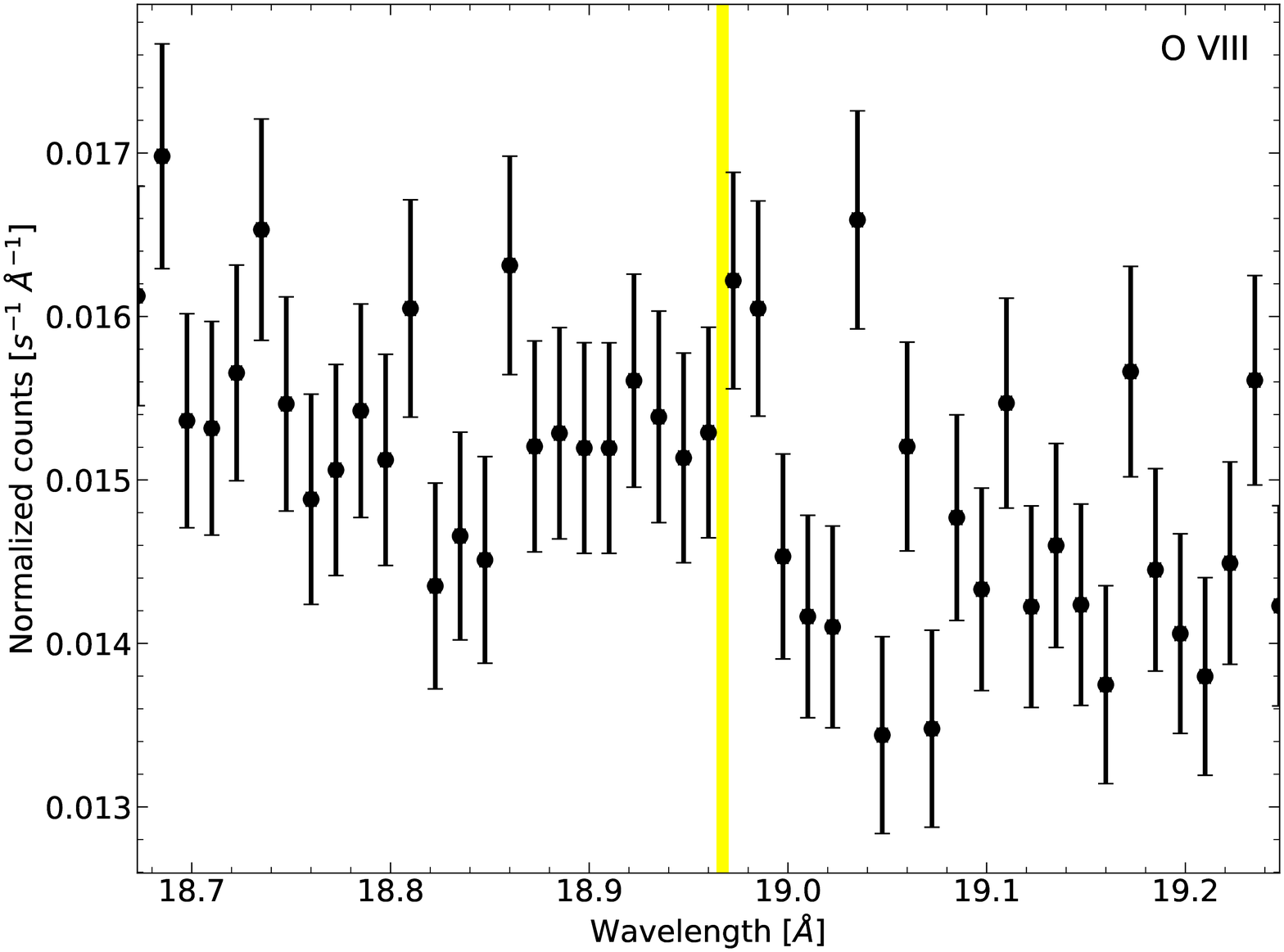}
	\vspace{0cm}
	\caption{Same as Figure \ref{fig:h1821oline}, but the spectra are shown at the rest-frame wavelength of Ne\,IX, Ne\,X, N\,VI, N\,VII, and O\,VIII ions. At the wavelength of these metal lines, we do not detect statistically significant absorption features, but do calculate upper limits.} 
	\label{fig:h1821lines}
\end{center}
\end{figure*}

\subsection{Spectral analysis} 
\label{sec:xspec}

In this work, we aim to probe whether the \textit{Chandra} LETG spectrum of H\,1821+643 exhibits X-ray absorption lines originating from the most abundant ions. Specifically, we investigate the N\,VI, N\,VII, O\,VII, O\,VIII, Ne\,IX, and Ne\,X ions, which are the most abundant ions in the WHIM and the LETG bandpass. Although we may also expect absorption lines at the wavelengths of C\,V and C\,VI, at the (redshifted) wavelength of these lines the effective area of ACIS-LETG is about an order of magnitude lower than that at the wavelength of the (redshifted) O\,VII line. Therefore, it is virtually impossible to probe the existence of  C\,V and C\,VI absorption lines in the stacked spectrum of H\,1821+643 using ACIS-LETG observations.

Before we investigate the stacked dataset, we probe whether the blueshifted but unstacked X-ray spectra exhibit absorption lines. We find that none of the individual spectra show statistically significant absorption lines. This implies that our stacked spectrum is not influenced by a single absorbing system. In the next step, we study the X-ray spectrum that was stacked based on the redshift of UV absorption line systems (Section \ref{sec:uv}).

\begin{table*}[!]
\caption{List of stacked equivalent widths and column densities with upper limits for non-detections}
\begin{minipage}{18cm}
\renewcommand{\arraystretch}{1.3}
\centering
\begin{tabular}{l  c c c c c c}
\hline
\hline
Metal line & O\,VII & O\,VIII & Ne\,IX & Ne\,X & N\,VI & N\,VII \\
\hline
Wavelength [$\rm \AA$] & 21.602 & 18.967 & 13.447 & 12.131 & 28.787 & 24.771 \\
Equivalent width [$\rm{m \AA}]$ & $4.1\pm1.3$ &  $<0.7$ & $<0.7$ & $<0.8$ &  $< 0.6$ \\
Ion column density [$10^{15} \ \rm{cm^{-2}}$] & $1.4 \pm 0.4$ & $ <0.5$ & $<0.6$ & $<2.0$  &$ <0.3$  &$ <0.1$ \\
\hline
\end{tabular}
\end{minipage}
\label{tab:ew}
\end{table*}

In Figure \ref{fig:h1821oline}, we present the stacked spectrum of H\,1821+643 in the $21.3-21.9  \, \rm{\AA} $ wavelength range. Visual inspection of the spectrum reveals an absorption line feature at $\lambda \approx 21.60 \, \rm{\AA}$, which is consistent with the rest-frame wavelength of  the O\,VII ion ($\lambda_{\rm O\,VII} = 21.602 \, \rm{\AA}$). Specifically, we detect a feature with four bins below the continuum level, where the central and deepest bin is located at $\lambda \approx 21.60 \, \rm{\AA}$. In addition, the feature exhibits a symmetric, Gaussian profile, which is consistent with that of absorption lines. These arguments imply that we have detected an O\,VII absorption line in the stacked spectrum of H\,1821+643. 

To characterize the absorption line at the wavelength of the O\,VII ion and to probe the statistical significance of this feature, we carry out spectral fitting with \texttt{XSpec} \citep{1996ASPC..101...17A}. We utilize a two component model that consists of an absorbed power law model that represents the continuum of the background AGN and a Gaussian absorption line profile, which is an appropriate model for weak lines. During the fitting procedure, the only fixed parameter is the wavelength of the line, which was fixed at $\lambda = 21.602$\,$\rm \AA$. {The intrinsic line width was set to 0 and, along with other model parameters (line normalization, power law normalization, and photon index of power law), it was a free parameter of the fit.} To determine the detection significance, we use the normalization parameter of the Gaussian line profile and the corresponding uncertainties.

The spectral fitting confirms the existence of an absorption line at  $\lambda = 21.602 \, \rm{\AA}$. Based on the \texttt{XSpec} fitting, we establish that the statistical significance of the absorption line is  $ 3.3 \, \sigma$. We emphasize that this detection is \textit{not} the result of a \textit{blind} search. Indeed, in this work -- unlike in many previous studies --  we did not employ numerous redshift trials to identify a tentative absorption line. Instead, the redshifts of the Ly-$\rm{\alpha}$ absorption lines, which were used to perform the stacking analysis, were known a priori from UV studies. Therefore, the derived $ 3.3 \, \sigma$ statistical significance represents the actual confidence level of the detection. To confirm that the centroid of the line agrees with the rest-frame wavelength of the O\,VII ion, we repeated the fitting procedure, but allowed the line centroid to vary.  We obtained the best-fit line centroid of $\lambda_{\rm obs} = 21.593^{+0.013}_{-0.006} \, \rm{\AA}$, which is consistent with the wavelength of the O\,VII ion. Thus, for the first time, we detect a definite O\,VII absorption line in the stacked spectrum of H\,1821+643.  

For the O\,VII detection toward the sightline of H\,1821+643, we calculated the equivalent width of the line and obtained $EW_{\rm O\,VII} = (4.1\pm1.3)$\,m\AA. 
{Given that the source is optically thin, and thus the detected absorption line is unsaturated, we can convert $EW_{\rm O\,VII}$ to the column density of the O\,VII ion using the linear formula:}
\begin{equation}
\label{coldens}
\begin{split} N\mathrm{(O\,VII)} & = 3.48 \times 10^{15} \ \mathrm{cm}^{-2} \left(
\frac{EW_{\mathrm{O\,VII}}}{10\;
\mathrm{m\AA}}\right) \\ & = (1.4\pm0.4) \times 10^{15} \ \mathrm{cm^{-2}}
\end{split}
\end{equation} 
where $EW_{\mathrm{O\,VII}}$ is the rest-frame equivalent width. 
{We emphasize that the stacked O\,VII equivalent width, and thus the column density, represent average values, according to the definition of equivalent width ($EW \equiv {\mathrm{F_{line}}}/{\mathrm{F_{continuum}}}$).}

In Figure \ref{fig:h1821lines}, we present the stacked spectra at the rest-frame wavelength of Ne\,IX, Ne\,X, N\,VI, N\,VII, and O\,VIII, where we highlight  $0.6 \ \rm{\AA} $ regions around the expected position of these lines. The data do not reveal the existence of prominent  absorption lines at the wavelengths of these ions. We also carry out spectral fitting, which confirms the non-detection of these lines in the stacked spectrum of H\,1821+643. In the absence of detections, we computed $1\sigma$ upper limits on the equivalent widths and column densities of the other major metal lines. The results are  summarized in Table~\ref{tab:ew}.

\subsection{Monte Carlo simulations}
\label{sec:mc}

Based on the \texttt{XSpec} analysis of the stacked spectrum of H\,1821+643, we obtained a $3.3 \, \sigma$ detection of an absorption line at the rest-frame wavelength of the O\,VII ion (Section~\ref{sec:results}). To further confirm the statistical significance of this detection, we utilize  a complementary and independent technique. Specifically, we perform Monte Carlo simulations and assess the possibility that the detection is the result of a chance coincidence.

To carry out the simulations, we mimic the observed data by generating a set of 17 random redshifts in the range of $z=0-0.297$, where the upper limit corresponds to the redshift of H\,1821+643. Using the random set of redshifts, we analyze the spectrum of H\,1821+643 following the steps described in Section~\ref{sec:data}. Specifically, we corrected for overlapping redshifts, we blueshifted the spectra and response files to the rest-frame wavelength, and we stacked the spectra. We fit the  stacked spectrum with  \texttt{XSpec} following Section \ref{sec:xspec} and we determined the detection significance of a potential absorption line feature at $21.602 \ \rm{\AA}$. To obtain statistically meaningful results, we generated $10^4$ random redshift sets (each containing 17 redshifts) and carried out this procedure for every randomly generated set of spectra.

\begin{figure}[t]
\begin{center}
\leavevmode
	\epsfxsize=0.48\textwidth \epsfbox{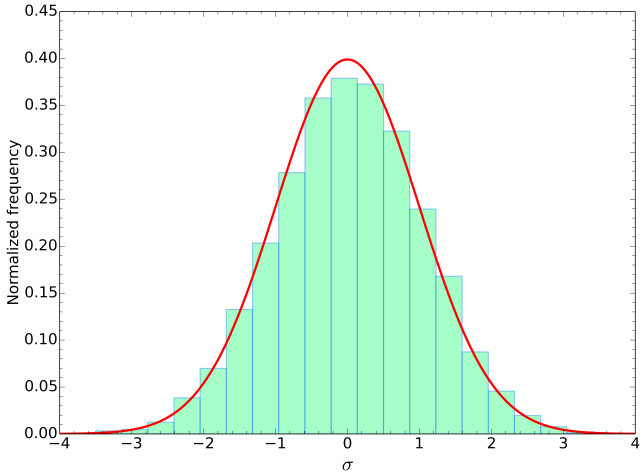}
	\caption{The distribution of statistical significances obtained from stacking and fitting (at $\lambda = 21.602 \rm \AA$) $10^4$ redshift sets, {with each set including $\leq17$ overlap-corrected random redshifts.} The overplotted curve represents the standard normal distribution. Overall, we obtain 3 random redshifts sets showing a $>3.3\sigma$ absorption line, which is consistent with the expectation.}
\label{fig:mc}
\end{center}
\end{figure}

In Figure \ref{fig:mc}, we present the distribution of the statistical significances of a possible line for the $10^4$ random redshift sets. The observed distribution is well described with a standard normal distribution. After fitting the randomly stacked spectra, we find that three random redshift sets show an absorption feature with $\geq3.3 \, \sigma$ significance and four random redshift sets exhibit an emission feature with $\geq3.3 \, \sigma$ significance. Hence, the chance coincidence of detecting a $\geq3.3 \, \sigma$ absorption line is $3\times10^{-4}$. This result is fully consistent with the detection significance obtained from the  \texttt{XSpec} fitting. Thus, our Monte Carlo simulations independently demonstrate that the detection of the O\,VII absorption line in the stacked spectrum of H\,1821+643 is at the $3.3\, \sigma$ confidence level.

To probe whether the randomly stacked spectra exhibit the characteristics of typical absorption or emission lines, we visually inspected the 7 spectra that exhibit a $\geq3.3\sigma$ detection. Typically, a prototypical Gaussian absorption/emission line has a symmetric profile and the lowest/highest data points are coincident with the rest-frame wavelength of the O\,VII ion. As discussed in Section \ref{sec:xspec}, the O\,VII absorption line detected in the stacked spectrum of H\,1821+643 exhibits these characteristic (Figure \ref{fig:h1821oline}). However, only one out of the seven random spectra with $\geq3.3\sigma$ significance, shows these characteristics. Hence, the random occurrence rate of an absorption/emission line with $\geq3.3 \, \sigma$ statistical significance with a Gaussian profile is $10^{-4}$. Assuming standard normal distribution, this chance coincidence corresponds to a statistical significance of $3.9 \, \sigma$.

\subsection{Verifying the stacking method}
\label{sec:verifying}

It is necessary to verify our stacking analysis is  because individual (i.e. unstacked) spectral features are not detectable, hence visual confirmation of the stacking is not possible. To this end, we used two approaches to probe the accuracy of our technique, while assuming an equal contribution from individual spectral lines.

First, we confronted the single and co-added spectra and we then examined them from bin to bin. This demonstrates that the stacked spectrum is the sum of the individual spectra, thereby confirming  the applicability of our analysis.

Second, as a more extensive approach, we repeated our stacking analysis on simulated spectra containing \textit{undetectable} (i.e. low statistical significance) absorption features, similar to the real data, and we then examined the line detection significance of the simulated, stacked spectra. 

To simulate the observations, we used the \texttt{Xspec} command \texttt{fakeit}, which allows us to create spectra using the model and instrumental response files of the observed data. The two important parameters of the applied model, containing a power law and a Gaussian line profile (Section \ref{sec:xspec}) are the normalization and the wavelength of the absorption line. We adjusted the line normalization in correspondence with {the detected} absorption feature. According to the number of foreground systems in the sightline of H\,1821+643, we defined 17 evenly spaced redshifts within the redshift of the AGN. We then used the corresponding wavelengths (with the O\,VII rest-frame wavelength as zero-point) as the location of the simulated absorption features. For the remaining model parameters, we adopted the values obtained from the fitting of the real unstacked data and we also used the unstacked value of 470\,ks for the exposure time. This way, we reproduced the spectrum of H\,1821+643 17 times, each with a low statistical significance absorption feature at one of the 17 \textit{fake} wavelengths.

Because of the random nature of \texttt{fakeit} simulations, the statistical significance of the stacked spectral feature is expected to follow a normal distribution. If our stacking analysis is valid, then the distribution should peak at ${\sim} 3 \, \sigma$. To have a sufficient sample size, we repeated the \texttt{fakeit} simulation 1000 times, resulting in 17000 unstacked spectra (and 1000 stacked spectra) using the same redshift set for each run. These spectra were then stacked and fitted following the methid given in Sections \ref{sec:data} and \ref{sec:mc}. The detection significances of the stacked spectra resulted in a normal distribution with a mean value of $-2.8 \sigma$ and standard deviation of 1.1, which implies that our stacking analysis is valid. In addition, we also fit the unstacked spectra at the simulated wavelengths, which also produced a normal distribution {for the detection significances} with mean value of $-0.8 \sigma$ and standard deviation of 1.0, corresponding to individually undetectable absorption lines. This verifies the suitability of the simulated spectra.


\begin{figure*}[!]
\begin{center}
\leavevmode
	\epsfxsize=0.48\textwidth \epsfbox{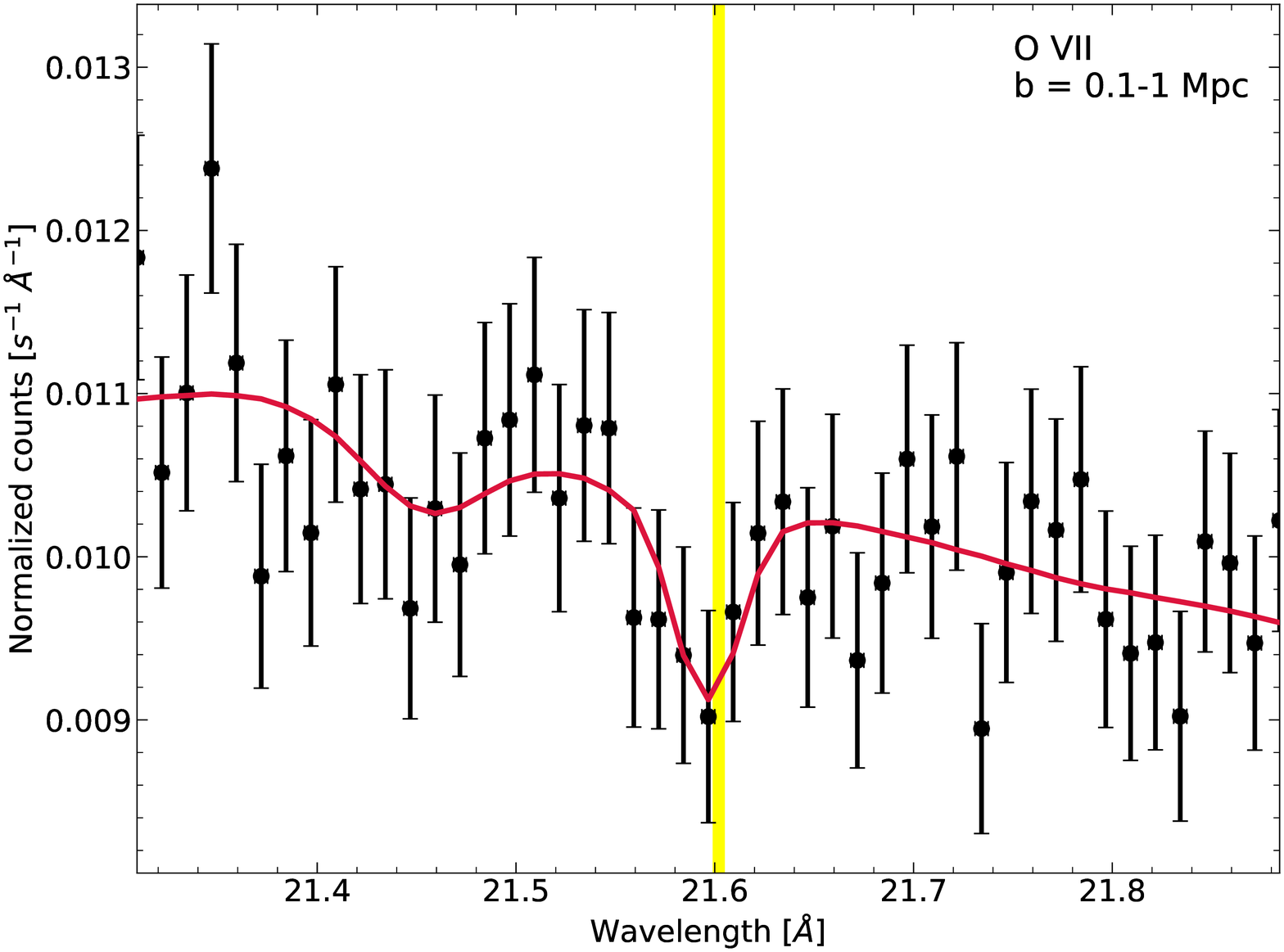}
	\epsfxsize=0.48\textwidth \epsfbox{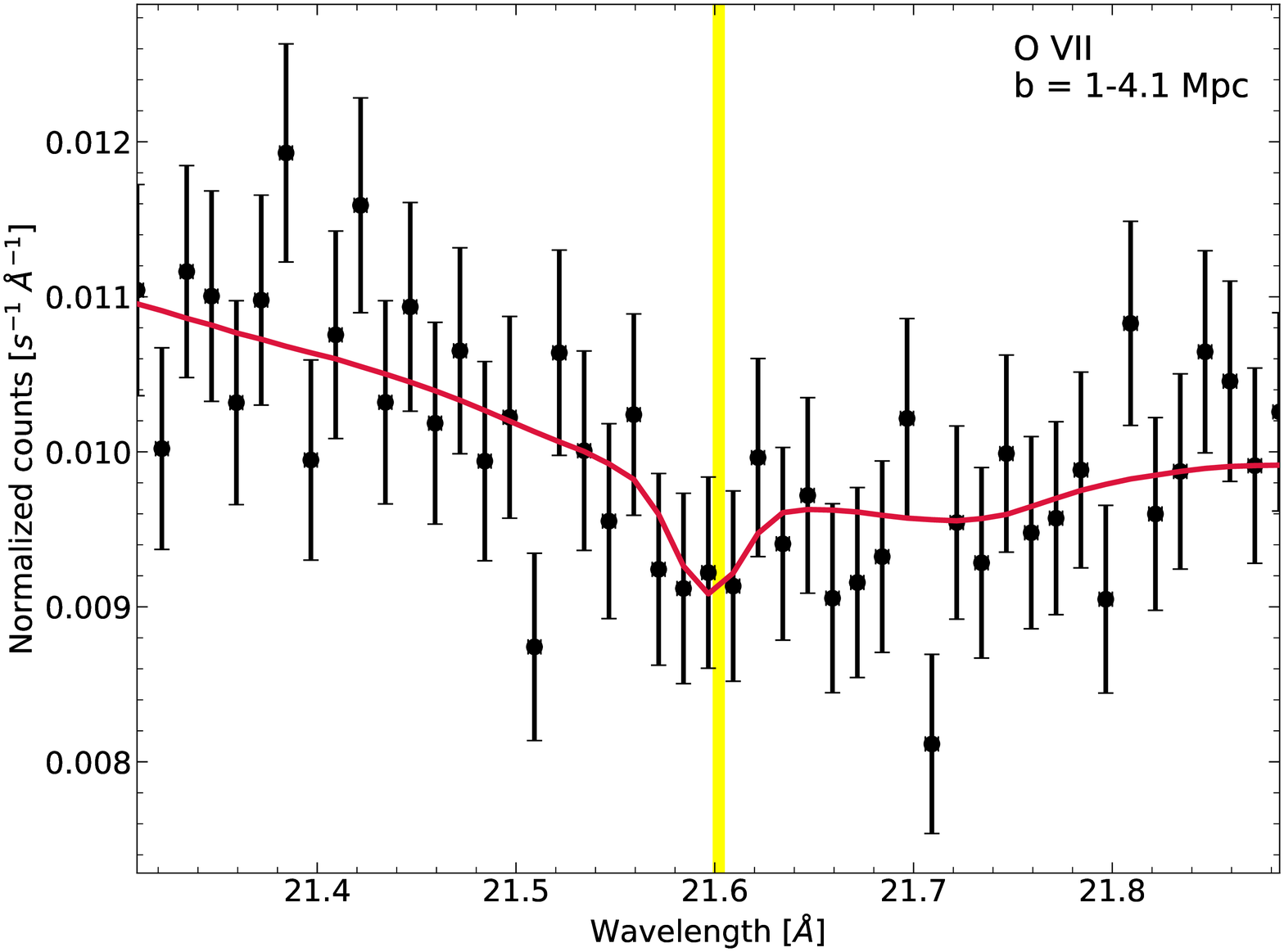}
	\vspace{0cm}
	\caption{The binned and stacked \textit{Chandra} ACIS-LETG spectra of H\,1821+643 around the wavelength of the O\,VII ion. The left-hand panel shows the stacked spectrum for absorption line systems whose impact parameter is $0.1<b<1$ Mpc, while the right-hand panel shows the stacked spectrum obtained for systems with $1<b<4.4$ Mpc. The detection significances are $2.9\sigma$ and $1.7\sigma$ for the low and high impact parameter systems, {respectively}.} 
	\label{fig:h1821bin}
\end{center}
\end{figure*}

\section{Discussion} 
\label{sec:disc}

\subsection{Cosmological mass density}
\label{sec:massdensity} 

The detection of the O\,VII ion in the stacked spectrum of H\,1821+643 allows us to derive the cosmological mass density of the O\,VII absorbers. Following the calculations carried out for UV absorption lines \citep[e.g.][]{2000ApJ...534L...1T}, {we express the OVII baryon density as }

\begin{equation}
\begin{split}
 \Omega_{\mathrm b} \mathrm{(O\,VII)} = &
 						\frac	{\mu m_{\mathrm p} H_{\mathrm 0}}		{\rho_{\mathrm c} c  } 
						\left[		\left(\frac{\mathrm{O}}{\mathrm{H}}\right) \, f_{\mathrm{O\,VII}} \, Z/Z_{\odot}		\right]^{-1} \cdot
\\ &						\cdot \frac	{\sum_i  N_i (\mathrm{O\,VII})}		{\Delta X}   
\end{split}
\end{equation}
\noindent
{where $\mu = 1.3$ is the mean atomic weight, $m_{\mathrm p}$ is the proton mass, $\rho_{\rm c}$  is the critical density of the universe, $({O/H}) = 4.94 \times 10^{-4} $ is the solar abundance of oxygen relative to hydrogen \citep{2009ARA&A..47..481A}, $f_{\rm{O\,VII}}$ is the O\,VII ionization fraction, $Z/Z_{\odot}$ is the metallicity of the absorber, $\sum_i  N_i (\rm{O\,VII})$ is the measured O\,VII column density summed up for $i=17$ absorption line systems, and $\Delta X$ is the absorption distance interval defined by \citet{1969ApJ...156L...7B}. } Assuming that all 17 absorption line systems contribute equally, the OVII baryon density is: 

\begin{equation}
 \Omega_{\mathrm b} \mathrm{(O\,VII)} = (0.0023 \pm 0.0007) [Z/Z_{\odot} \, f_{\mathrm{O\,VII}}]^{-1} 
\end{equation}
However, if fewer than 17 systems contribute, then this value corresponds to an upper limit. Hence the OVII baryon density can be expressed as: 

\begin{equation}
 \Omega_{\mathrm b} \mathrm{(O\,VII)} = \frac{i}{17} (0.0023 \pm 0.0007) [Z/Z_{\odot} \, f_{\mathrm{O\,VII}}]^{-1}  \ ,
\end{equation}
where $i$ is the number of absorbers contributing to the signal and also where $i$ must be more than a few because none of the individual systems are detectable (see Section~\ref{sec:xspec}).
\noindent
In this formula, the major sources of uncertainty are the unconstrained metallicity of the WHIM and the ionization fraction of O\,VII.

Cosmological simulations showed that prominent filaments with higher column densities have higher metallicities  \citep{2012ApJ...753...17C}. Specifically, for the column density of $N_{\rm O\,VII} = 1.4\times10^{15} \ \rm{cm^{-2}}$, the expected metallicity is $Z = 0.18^{+0.17}_{-0.09} \, Z_{\odot}$. 

Specifying the ionization fraction of different species in the WHIM is not straightforward. Generally, ionization comes from two distinct processes: collisional ionization and photoionization. Assuming only collisional ionization, O\,VII dominates the widest temperature range of $T \approx 10^{5.5} -  10^{6.3}$\,K, and at lower temperatures, O\,V has the highest ionization fraction, overwhelming the O\,VI ion. 
However, in the low-density WHIM, collisional ionization plays a minor role in the formation of O\,V--O\,VIII ions at temperatures between ${\sim}10^4-10^7$\,K but may be relevant at higher temperatures ($T>10^7$\,K) or hydrogen densities above $10^{-3}$\,cm\textsuperscript{-3} \citep{1998ApJ...509...56H}.
Therefore, at temperatures relevant for WHIM filaments, photoionization is the dominant process. Specifically, UV radiation is responsible for the ionization of lithium-like species (e.g. O\,VI), while X-ray radiation plays a role in the ionization of helium-like and hydrogen-like ions (e.g. O\,VII and O\,VIII). The effect of photoionization depends strongly on the intensity and the shape of the external radiation field, and the hydrogen number density of the WHIM \citep[e.g.][]{2003ApJ...594...42C,2006ApJ...650..573C,2006PASJ...58..657K}. 
However, assuming collisional and photoionization equilibrium is still an oversimplification.
For the O\,VII--O\,IX ions at densities of the WHIM, the timescales required to reach ionization equilibrium may be longer than the Hubble time \citep{2006PASJ...58..641Y}. While general ionization processes (i.e. collisional ionization and photoionization) can be modeled by photoionization codes, the importance of the latter effect requires hydrodynamical simulations. Other hydrodynamical effects, such as supernova feedback, gas shocks produced by AGN or star formation activity, merging or accretion processes, and the anisotropic temperatures occurring at sites of nonlinear structure formation, may also be significant sources of ionization and cannot be neglected. 
{Based on hydrodynamical simulations of \citet{2012ApJ...753...17C}, we assume a temperature of ${\sim} 10^{6}$\,K for a single WHIM filament in the sightline of H\,1821+643.}
Considering the large parameter space involved by the different ionizing mechanisms, we rely on non-equilibrium simulations from \citet{2005AAS...20717517J} and results of \citet{2003ApJ...582...82M} to estimate the ionization fraction of O\,VII, for which we adopt $f_{\mathrm{O\,VII}} = 0.75$.

\begin{figure*}[!]
\begin{center}
\leavevmode
	\epsfxsize=0.48\textwidth \epsfbox{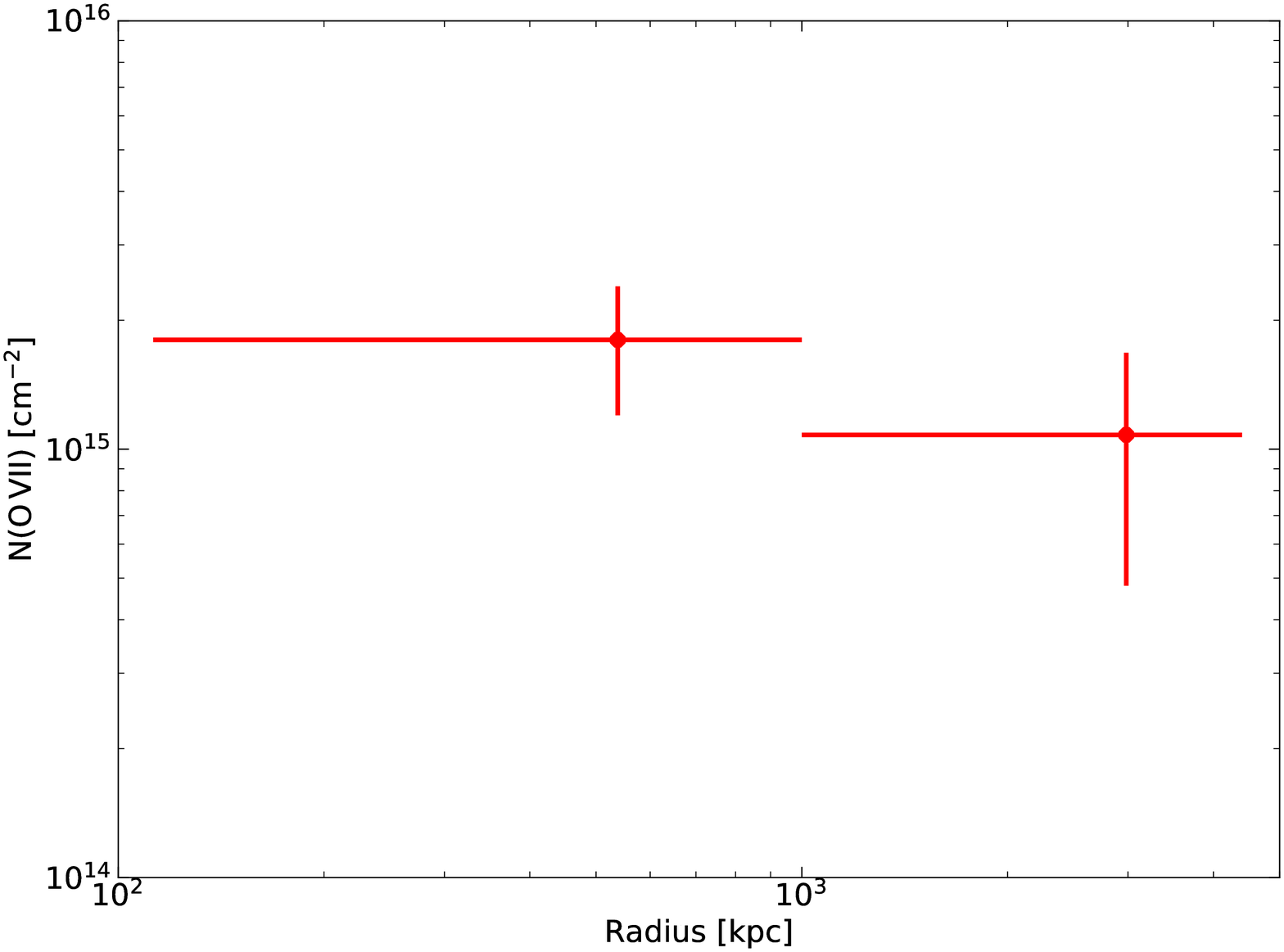}
	\epsfxsize=0.48\textwidth \epsfbox{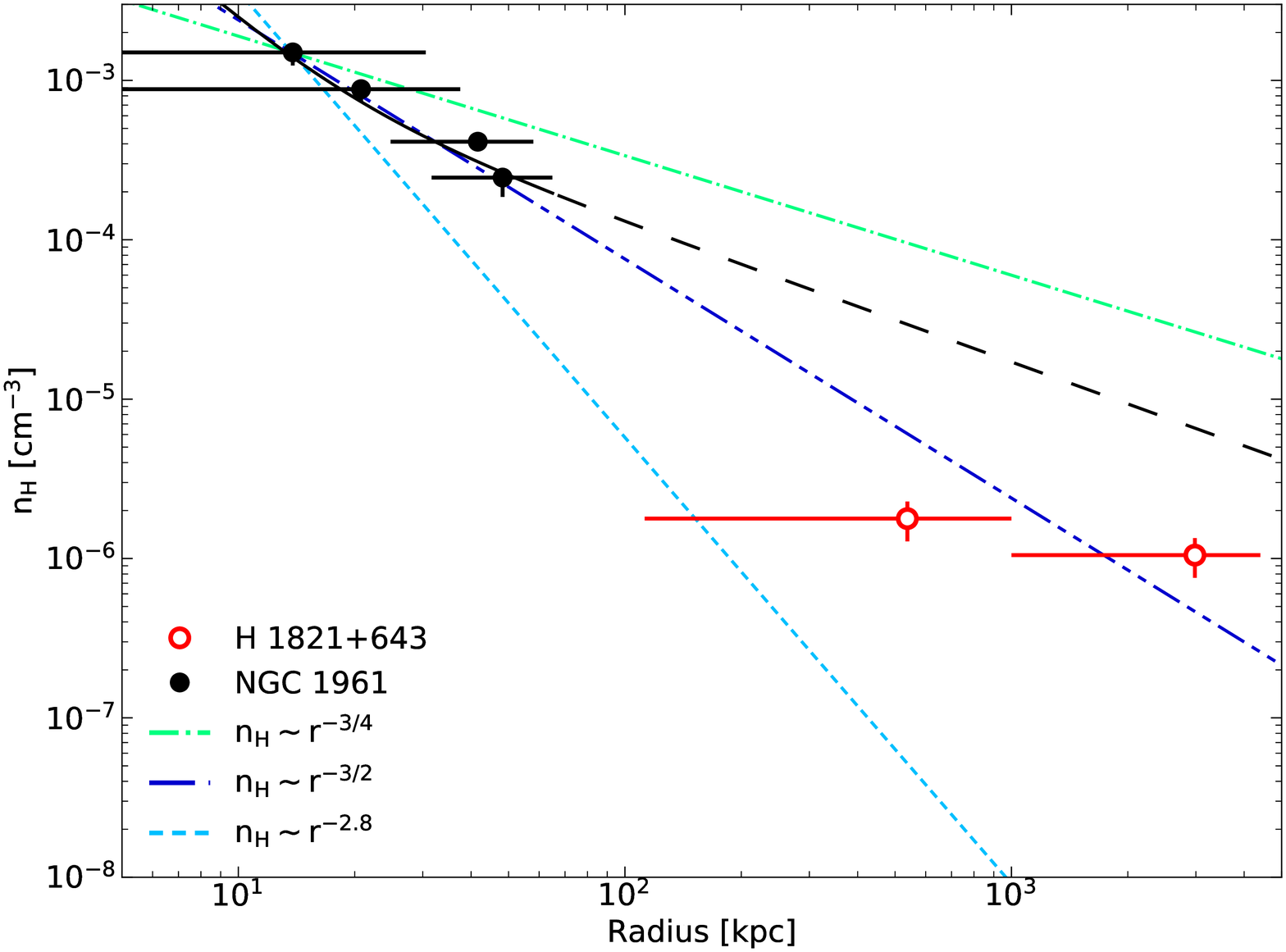}
	\caption{{Left}: O\,VII column density distribution of the hot gas as a function of impact parameter. To construct this plot, we binned the absorption line systems as a function of their impact parameter. We note that due to the fewer absorption line systems in each {subsample}, we achieve weak, $1.7\sigma$ and $2.9\sigma$, detections. Note that within statistical uncertainties the {O\,VII} column density distribution is invariant as a function of impact parameter.	
	{Right:} 
	Density distribution of hydrogen as a function of impact parameter. To derive the hydrogen densities we used Equation~\ref{odens}. For the path length we assumed 5 Mpc, which is the average width of the filaments. {The gas density in the WHIM (red empty circles) is $n_{\mathrm{H}} {\simeq} (1-2)\times 10^{-6} $\,cm\textsuperscript{-3} that corresponds to an overdensity of $\delta {\simeq} 5.3-8.9$, which values are typically expected for WHIM filaments.}
	The black filled circles represent the measured values for massive spiral, NGC\,1961, fitted with black solid curve, while black long-dashed line is extrapolation \citep{2015ApJ...804...72B,2016MNRAS.455..227A}. Note that the gas density in the outer halo of a massive spiral galaxy {is about two orders of magnitude higher than that in the WHIM filaments.} The green dotted-dashed line represents the case where almost all the baryons are located within $r_{200}$, while the two steepest curves -- one representing a fit to simulated halo distribution and the other representing a  collisionless NFW halo -- have a better agreement with our measurements \citep{2017SPIE10397E..0QS}.}
	\label{fig:densprof}
\end{center}
\end{figure*}

{Using $Z \approx 0.18 \,Z_{\odot}$ for metallicity and $f_{\mathrm{O\,VII}} =  0.75$ for the O\,VII ionization fraction with $H_{\rm 0} = 70 \ \rm{km \ s^{-1} \ Mpc^{-1}}$, and assuming that all 17 systems contribute equally to the detected signal, we estimate that the O\,VII baryon mass density is $\Omega_{\rm b} \rm(O\,VII) = 0.017 \pm 0.005$. 
Given that the cosmic baryon mass density is $\Omega_{\rm b} = 0.045$, we conclude that O\,VII absorbers contribute $(37.5 \pm 10.5)\%$ of the total baryon content of the Universe.
This is in good agreement with recent results from IllustrisTNG simulations, where the WHIM constitutes $\sim 47$\,\% of the baryons at $\mathrm{z} = 0$ \citep{2018arXiv181001883M}. }

\subsection{Binning the galaxy sample}
\label{sec:binning}
 
The galaxies in the vicinity of H\,1821+643 exhibit a wide range in impact parameters ($b = 0.1 - 4.4$\,Mpc), which covers regions from the outer halos of galaxies  (${\sim}0.1$ Mpc) to the scales of the WHIM (${\sim} 4.4$\,Mpc). To study the O\,VII absorption line as a function of impact parameter, we binned the absorption line systems as a function of their impact parameter. Specifically, we divided the absorption line systems into two {subsamples}. The first {subsample} includes eight galaxies with $0.1\,\mathrm{Mpc}<b<1$\,Mpc, while the second {subsample} includes nine galaxies with $1\,\mathrm{Mpc}<b<4.4$\,Mpc. We emphasize that the impact parameter exceeds the virial radius for almost all foreground galaxies (Table \ref{tab:abs}), except for two galaxies whose virial radius exceeds the impact parameter. However, both of these galaxies have relatively low halo mass ($M_{\rm halo}  \lesssim 7\times10^{11} \ \rm{M_{\odot}}$), implying that these galaxies are not likely to host extended X-ray halos around them. Thus, both the inner and outer {subsample} primarily probe the large-scale WHIM filaments. 

Following the procedure outlined in Section \ref{sec:data}, we stack the absorption line systems in each {subsample} and fit the O\,VII absorption line with a Gaussian absorption model as described in Section~\ref{sec:xspec}. Given the fewer absorption line systems in the {subsamples}, we obtain weak detections in the inner and outer {subsamples} at the $2.9\sigma$ and $1.7\sigma$ level, respectively. We present the spectra and the best-fit models in Figure \ref{fig:h1821bin}.
{Taking these signals as detections, we derive the {O\,VII} equivalent width and column density,  and obtain $EW_{\rm in} = (5.2\pm 1.6) \ \rm{m\AA}$ and $EW_{\rm out} = (3.1\pm 1.8) \ \rm{m\AA}$ for the {subsamples with smaller and larger impact parameters, respectively.} 
These values correspond to O\,VII column densities of $N(\mathrm{O\,VII})_{\rm in}=(1.8 \pm 0.6) \times 10^{15}$\,cm\textsuperscript{-2} and $N(\mathrm{O\,VII})_{\rm out}=(1.1 \pm 0.6) \times 10^{15}$\,cm\textsuperscript{-2}. }

In the left-hand panel of Figure \ref{fig:densprof}, we display the column density as a function of impact parameter. Within statistical uncertainties, we find that the column density is invariant as a function of impact parameter. While a decreasing trend may exist, the present data is not sufficient to better constrain the dependence of column density as a function of impact parameter.

In the right-hand panel of Figure \ref{fig:densprof}, we depict the hydrogen gas density distribution as a function of impact parameter. We deduced the $n_{\mathrm{H}}$ hydrogen density for the inner and outer {subsamples} using:
\begin{equation}
\label{odens}
\begin{split}
n_{\mathrm{H}} & = \frac{N(\mathrm{O\,VII})} { f_\mathrm{O\,VII} \
Z/Z_\sun  \ ({O}/{H}) \ r} \\
\end{split}
\end{equation}
where $N(\mathrm{O\,VII})$ is the {average O\,VII} column density of a single WHIM filament, $Z/Z_\sun \approx 0.18$ is the metallicity of the gas (Section~\ref{sec:massdensity}), $({O/H}) = 4.94 \times 10^{-4} $ is the solar abundance of oxygen relative to hydrogen \citep{2009ARA&A..47..481A}, $f_\mathrm{O\,VII} = 0.75$ is the ionization fraction of O\,VII (Section~\ref{sec:massdensity}), and $r$ is the path length. Given that the observed signal represents the large-scale WHIM, we assumed $r = 5$ Mpc, which is the typical width of WHIM filaments \citep{2014MNRAS.441.2923C}.
{We note that filaments have complex shape and structure, and they also vary in length and width, depending on their environment \citep{2006MNRAS.370..656D}.}
{We find that the gas density of a WHIM filament is $n_{\mathrm{H}} \simeq (1-2) \times 10^{-6} $\,cm\textsuperscript{-3},
which corresponds to overdensities of $\delta \simeq 5-9$. These values are typically expected for WHIM filaments \citep{2007MNRAS.381...41H,2010MNRAS.408.2163A,2014MNRAS.441.2923C}. Thus, our detections are consistent with a picture in which the observed absorption lines originate from large-scale WHIM filaments.}

In the right-hand panel of  Figure \ref{fig:densprof}, we also show the gas density profile for the massive spiral galaxy, NGC\,1961. Emission studies have explored the gas density profile to about 60\,kpc radius around the galaxy (the CGM), which corresponds to about ${\sim}15\%$ of the virial radius \citep{2015ApJ...804...72B,2016MNRAS.455..227A}. Combining these data with our absorption line study allows us to construct, for the first time, a gas density profile that simultaneously explores the outer scales of the CGM and the WHIM. 
We also adopted and display three model curves from \citet{2017SPIE10397E..0QS}\footnote{See also \texttt{http://www.arcusxray.org/SPIE2017-Smith-Arcus-sm.pdf}}. The steepest curve ($n_{\mathrm{H}} \sim r^{-2.8}$) represents a Navarro--Frenk--White (NFW) halo, assuming collisionless particles. The curve with the lowest slope ($n_{\mathrm{H}} \sim r^{-3/4}$) models a mass distribution where most of the baryons reside within $r_{200}$. Finally, the $n_{\mathrm{H}} \sim r^{-3/2}$ curve is fitted to a simulated galactic halo distribution.
When comparing the observations,
we find that the average density of the WHIM filaments {is about two orders of magnitude lower} and is also significantly lower than predicted from the extrapolated density profile established for
NGC\,1961.
A better agreement is obtained with the $n_{\mathrm{H}} \sim r^{-2.8}$ or $n_{\mathrm{H}} \sim r^{-3/2}$ curves, within the {statistical} uncertainties. Although this measurement is insufficient to determine an exact model for circumgalactic regions, it excludes predictions with higher densities.

\section{Conclusions}
\label{sec:conclusion}

In this work we report the detection of O\,VII ion absorption from intervening absorption line systems along the sightline of the X-ray bright AGN, H\,1821+643. Our results can be summarized as follows: 

\begin{itemize}
\item We utilize 470 ks \textit{Chandra} ACIS-LETG observations of H\,1821+643 and we stack the data according to the redshifts of 17 previously detected UV absorption line systems. The equivalent exposure time of the data is $8.0$\,Ms, which allows us to probe unprecedentedly low column densities.

\item After blueshifting and stacking the spectrum, we detect an absorption line with $3.3 \, \sigma$ statistical significance at the wavelength of the O\,VII ion. Monte Carlo random simulations suggest that the likelihood of chance coincidence is $7\times10^{-4}$, which is consistent with the detection significance measured through \texttt{XSpec} fitting.

\item {We find that the equivalent width of the stacked O\,VII line is $(4.1\pm1.3)  \ \rm{m\AA}$ and the {corresponding O\,VII} column density is $(1.4 \pm 0.4) \times10^{15} \ \rm{cm^{-2}} $. These values represent an average WHIM filament in the sightline of H\,1821+643.}

\item We do not detect absorption lines at the wavelength of other major helium-like or hydrogen-like ions, such as Ne\,IX, Ne\,X, N\,VI, and N\,VII, {which was expected for these weaker lines}. 

\item {Based on the stacked data, we address the global missing baryon problem and compute the O\,VII cosmological mass density of $ \Omega_{\mathrm b} \mathrm{(O\,VII)} = (0.0023 \pm 0.0007) [Z/Z_{\odot} \, f_{\mathrm{O\,VII}}]^{-1}$, assuming that all 17 systems contribute equally. We find that missing baryons likely reside in the form of tenuous hot gas in the WHIM.}

\item {We establish the O\,VII column density and hydrogen density profile of the hot gas, and we constrain the absorbing gas to have an overdensity of $\delta \simeq 5-9$, which is typically expected for WHIM filaments.}

\end{itemize}

\bigskip

\begin{small}
\noindent
\textit{Acknowledgements.}{ We thank the referee for the constructive report, which greatly improved the manuscript.} We thank Doug Burke and the CIAO team for their help with the data analysis. This research has made use of Chandra data provided by the Chandra X-ray Center (CXC). This paper makes use of software provided by the CXC in the application package CIAO. The NASA/IPAC Extragalactic Database (NED) has been used in this work. Funding for the Sloan Digital Sky Survey IV has been provided by the Alfred P. Sloan Foundation, the US Department of Energy Office of Science, and the Participating Institutions. SDSS acknowledges support and resources from the Center for High-Performance Computing at the University of Utah. The SDSS website is www.sdss.org.  \'A.B., R.P.K., and W.R.F. acknowledge support from the Smithsonian Institute.

\end{small}

\bibliographystyle{apj}
\bibliography{paper4}

\end{document}